# Super Damping of Mechanical Vibrations

Ka Yan Au Yeung, Brian Yang, Liang Sun, Kehang Bai, Z. Yang*

Department of Physics, The Hong Kong University of Science and Technology, Clearwater Bay, Kowloon, Hong Kong, The People's Republic of China

**Abstract**

We report the phenomenon of coherent super decay (CSD), where a linear sum of several damped oscillators can collectively decay much faster than the individual ones in the first stage, followed by stagnating ones after more than 95 % of the energy has already been dissipated. The parameters of the damped oscillators for CSD are determined by the process of response function decomposition, which is to use several slow decay response functions to approximate the response function of a fast decay reference resonator. Evidence established in experiments and in finite element simulations not only strongly supported the numerical investigations, but also uncovered an unexplored region of the tuned mass damper (TMD) parameter space where TMD's with total mass less than 0.2 % of a stainless steel plate can damp its first resonance up to a damping ratio of 4.6 %. Our findings also shed light onto the intriguing underline connections between complex functions with different singular points.

**Introduction**

Damped vibrations are common phenomena in many physical systems, including merging black holes [1] and neutron stars [2]. The classical vibrations can be described by a simple expression in time domain and in frequency domain [3]. Mitigating unwanted vibration in machinery, buildings, bridges, airplane, satellites, etc. is still a great challenge in engineering science despite of worldwide intensive efforts over the last century [4, 5]. The distribution of vibration energy is mostly within the first few resonant modes of the structures. Since its invention nearly 100 years ago [6], the toned mass dampers (TMD) or dynamic vibration absorbers (neutralizers) are particularly effective in suppressing vibrations with discrete frequencies, because their maximum effect is within a narrow band width around their individual working frequency, which can be designed and tuned to suit particular application needs. As a result, they have been widely used to specifically target at the first few resonant modes in many structures [7 – 33]. For comprehensive summary and reviews, see [7 – 10]. A TMD can be generically described by a damped harmonic oscillator [6]. The first theoretical investigation of the optimum TMD design was carried out in 1928 [11, 12]. Optimizations with generic TMD's soon followed and are still being pursued today [11 – 29]. Most TMD's are made of cantilevers or mass-spring that each weighs about 100 g or more for working frequency below 100 Hz [30 – 32]. The optimizations have been limited to the parameter space where the total mass of the TMD's was a few percent of the primary structures, even though in some cases the TMD mass was up to 35 % of the primary structures [27, 28]. Damping the vibration in free bodies, such as satellites and airplanes, are even more challenging. While part of the vibration energy of a supported structure can be dissipated by the supporting bodies, i. e., the vibration energy of a building or bridge can partially be transmitted to the ground, or the vibration of a floor can be partially transmitted to the building, the vibration energy of a free body can only be dissipated internally. As a result, active control is the preferred choice for free bodies [33 – 42].

In this paper we report solid experimental evidence and theoretical analysis of super damping of the first resonant mode at 100 Hz of a nearly free elastic body (a stainless steel plate about 6.4 kg in mass) brought by effective TMD's with total mass less than 0.2 % of the plate. Under an impulse excitation, the decay of the first mode of the plate was nearly complete within a time scale of 100 ms, or about 10 periods of the oscillation. As the elastic body had negligible internal loss (Q > 3000), almost all the impulse energy was dissipated by the TMD's within 10 periods. This shows that all the previous studies have missed an important parameter space of TMD's with mass about 50 times lighter than the conventional ones. Such phenomenon is explained with the concept of coherent super decay. A design strategy of damper parameters based on response function decomposition is developed to achieve super decay.

**Basics of Coherent Super Decay**

Consider a group of five damped oscillators with combined time dependent displacement in the form

$$x(t) = \sum_{n=1}^{5} A_n e^{-\omega_n^I t} \cos(\omega_n^R t) \qquad (1),$$

where $\omega_n^I = \frac{\omega_n}{2Q_n}$, and $\omega_n^R = \omega_n \sqrt{1 - \frac{1}{Q_n^2}}$.

As a typical example for illustration, we choose $Q_n = 20$, $\omega_n = 1 + \delta\omega_n$. The other parameters are $\{A_n\}$ = {0.0515, 0.1516, 0.358, 0.229, 0.0792}, $\{\delta\omega_n\}$ = {− 0.09, − 0.0389, − 0.0048, 0.0267, 0.0792}. The decay curve of a representative oscillator, which is the $n = 3$ term in Eq. (1), is shown in Fig. 1(a) as the purple curve. Its decay time constant (DTC) is almost the same as an oscillator with angular frequency 1 Hz and quality factor (Q-factor) $Q = 20$. As the Q-factors of the other four oscillators are also equal to 20, and their resonant frequencies deviate from the central frequency of 1 Hz by at most 9 %, the DTC of each individual oscillator deviates from each other by at most 10 %. Intuitively, one would expect the DTC of the combined displacement given in Eq. (1) to be about the same as the $n = 3$ term, i. e., it decays like an oscillator with $\omega_0 = 1$ and $Q = 20$. After all, if all $\delta\omega_n = 0$, the combined decay would almost exactly follow the purple curve. The actual curve obtained by simple numerical computation using Eq. (1) and the parameters given above, however, is the green solid curve (middle) in Fig. 1(a), which decays almost twice as fast as any of the individual oscillators. In fact, the decay curve resembles that of an oscillator with $\omega_0 = 1\,\text{Hz}$ and $Q = 10$, which is depicted as the red (bottom) curve in Fig. 1(a). In other words, when the parameters $\{A_n\}$ and $\{\delta\omega_n\}$ are properly selected, the combined decay of several oscillators can be much faster than each individual ones. We refer to such phenomenon as the *coherent super decay* (CSD) of harmonic oscillators.

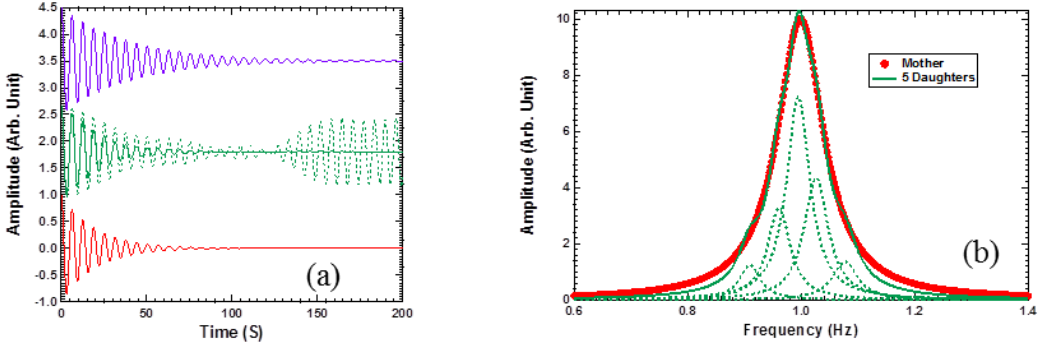

Figure 1, (a) Displacement as a function of time of several damped oscillators obtained by numerical computations. (b) The reference response function (red disks) and the approximation response function (solid green curve) made by the response functions of five individual oscillators (green dashed curve).

The CSD is caused by the coherent interference among the individual oscillators. To see this more clearly, the time dependent displacement without the damping factors $e^{-\omega_n^I t}$ in Eq. (1) was also calculated and shown as the green dashed curve in Fig. 1(a). It can be seen that the first minimum of the displacement by interference alone occurs at almost the same time when the combined decay (green solid curve) is near completion. At this moment, the phases of the individual oscillators are mutually destructive. Without damping, however, the displacement recovers its strength when the phases become mutually enhancing again as time goes by. With damping, sufficient vibration strength of each oscillator has already been consumed and the resurgence is negligible.

We now show an approach to determine the parameters that can realize CSD. According to classical mechanics [3], the time dependent displacement given in Eq. (1) is resulted from the frequency response function given below,

$$X(\omega) = \sum_{n=1}^{5} \frac{A_n}{\omega_n^2 - \omega^2 + i\omega\omega_n / Q_n} \qquad (2).$$

This is a typical frequency response function of a vibrating system with 5 degrees of freedom being excited. The time dependent displacement in response to a unit impulse $\delta(t)$ is given by

$$x(t) = \mathrm{Re}\left( \frac{1}{2\pi} \int_{-\infty}^{\infty} X(\omega) e^{i\omega t} d\omega \right) \qquad (3).$$

The integral can be calculated using standard path integral and residue theorem in complex analysis [41]. As the CSD behaves like a reference oscillator with resonant frequency = 1 and $Q_0$ = 10, we plot the imaginary part of the response function of the reference oscillator in the form $X_0(\omega) = \dfrac{1}{\omega_0^2 - \omega^2 + i\omega\omega_0/Q_0}$ in Fig. 1(b) as the red curve. The real part is not shown due to space limit. The green points are the imaginary part of the response function in Eq. (2) using the CSD parameters. It is no coincidence that the two curves are nearly the same, because the amplitudes $\{A_n\}$ and the resonant frequencies $\{\delta\omega_n\}$ were deliberately chosen to approximate, or mimic, the reference oscillator response function with $Q_0 = 10$. The green dashed curves are the response functions of the five mimicking resonators plotted individually. What has been shown here is what we refer to as *response function decomposition* (RFD), that is, if a response function with large line width can be mimicked by the sum of a number of response functions with much smaller line width, then the time dependent displacement generated by the two response functions will be almost the same, even though each mimicking oscillator decays much slower. That is the underline strategy to identify the amplitude and the central frequency parameters for the response functions with slow individual decay to collectively generate a much faster decay.

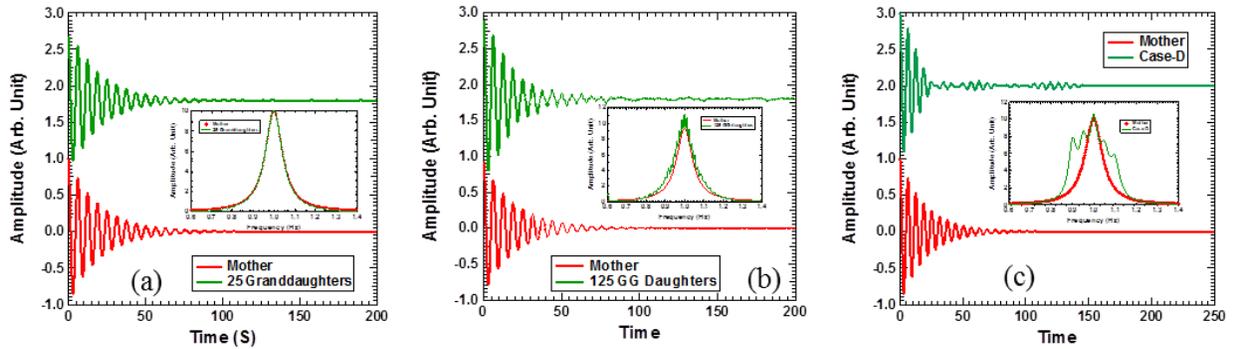

Figure 2, (a) The computed time dependent displacement of case-B (green curve) together with the reference (red curve). The insert depicts the response function of case-B (green curve) and that of the reference (red curve). (b) The corresponding quantities for case-C. (c) The corresponding quantities for case-D.

In the above example, we decompose one reference (mother) resonator of $Q_0 = 10$ into five $Q = 20$ daughter resonators (Case-A). The same can be done to decompose each daughter resonator into five granddaughters with $Q = 40$, so as to mimic the mother resonator with 25 granddaughter resonators, each with four times the Q-factor of the mother. The resulting time dependent displacement curve is shown in Fig. 2(a), together with the response function in the insert (Case-B). The corresponding results by 125 great granddaughter response functions with $Q = 80$ (Case-C) are shown in Fig. 2(b). In Case-C we deliberately kept the mimicking curve relatively rough while keeping the average line shape nearly tracing the original mother resonator. Despite the rough mimicking, the time dependent displacement curve of the 125 great granddaughters is still very close to that of the mother resonator.

If the reference oscillator response function is poorly mimicked (Case-D), as is shown in the insert of Fig. 2(c), the resulting decay will resurge after initial fast decay, as is shown in the main graph, and the decay lasts for quite a long time.

For more in-depth examination of the decay behavior, we calculate the quantity that is proportional to the remaining total energy of a vibrating elastic body, given by

$$E(t) = \int_t^\infty |x(t')|^2 \, dt' \qquad (4).$$

The curve by the mother (reference) resonator is shown in Fig. 3 as the red curve, which takes the simple form of $e^{-0.1t}$. We refer to such decay as a uniform decay, in that the DTC remains the same throughout the decay process. The corresponding curve by the five daughters (Case-A) is shown as the green curve in the same figure, together with the one made by the 25 granddaughters (Case-B, blue curve) and the 125 great granddaughter resonators (Case-C) as the purple curve. All of them exhibit similar decay in the first stage as the reference, and staged decays with different DTC's in different stages afterwards. This is the consequence that their mimicking of the uniform decay reference resonator is only approximate.

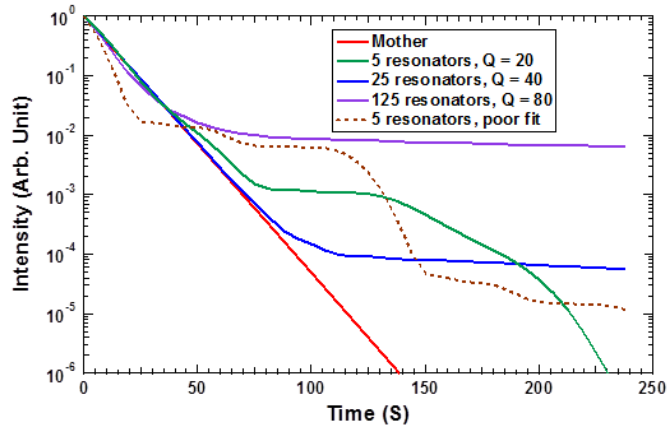

Figure 3, The remaining vibration intensity as a function of time for the cases studied above, together with that of the reference (red line).

For Case-A, the decay curve follows closely that of the reference resonator up to $t = 38$ s (6.05 periods for an angular frequency of 1 Hz), where $E = 2.56\,\%$, in the first stage decay. The quick drop levels off at around $t = 81$ s, after which the system is in the stagnating second decay stage. However, by then only 0.12% of the original vibration energy still remains, so the stagnation will not have any practical implications in vibration damping of a primary structure. The reference resonator will take 69 s to reach such energy reduction. Therefore, for all practical purposes in engineering Case-A can be regarded as decaying as fast as the reference.

The decay curve of Case-B (blue curve) matches well the reference up to $t = 69$ s, where $E = 0.1$ %. The curve for Case-C (purple curve) decays even slightly faster than the reference up to $t = 38$ s, where $E = 2.5$ %. This is consistent with its response function being a little bit wider in combined line width than the reference. It levels off at $t = 69$ s, with $E = 1.0$ %.

Case-D with poorly mimicking response function is also interesting. Its apparent line width is almost twice of that of the reference. So the question is at what time it will level off after the first stage fast decay. The result, shown as the brawn dashed curve in Fig. 3, indicates that it takes only $t = 25$ s (4 periods) to bring $E$ down to 1.7 %, as compared to 44 s for the reference. At the time of leveling off, only 2 % of the initial energy is still present. Therefore, in practical applications, the decay can already be regarded as complete, and the damper parameters chosen to create such response function are really good for practical optimization purposes.

**Experimental Verifications**

The CSD realized by following the RFD strategy can lead to the designs and realization of fast vibration decay of primary structures with dozens times lighter multiple TMD's than the ones reported in the literature [4 – 29], as will be shown in the experiments below.

The primary structure for the damping experiments is a 40 cm × 40 cm × 5 mm stainless steel plate (6.24 kg in mass) with free edges. Its first eigenmode is at 100.24 Hz with a Q-factor over 3000 (damping ratio $\eta = 0.5/Q = 1.7 \times 10^{-4}$). The plate was hanging by soft threads through the two clear holes near its top edge at about 10 cm apart. For frequency response measurements, the plate at the position 7 cm from the lower right corner along the plate diagonal line was attached to a shaker similar to the one reported earlier [42]. The response displacement was measured by a miniature accelerometer at the corresponding position relative to the left lower corner. Single frequency excitation was used, and the excitation and the response displacement signals were measured by lock-in amplifiers. For free vibration decay measurements, the shaker was removed, and a gentle knock was applied near the same position where the shaker was originally attached, and the time dependent displacement of the plate vibration was recorded by a data collection card controlled.

The TMD's used in this study were made of miniature decorated membrane resonators (DMR's) reported earlier [42, 43]. The diameter of each stretched membrane was 25 mm with its boundary fixed on the circular plastic frame about 5 mm in width and 5 mm in height. The thickness of the rubber membrane is 0.2 mm. The frame weighs about 0.5 g. A steel platelet 10 mm in diameter and 1.0 g in mass was attached to the center of the membrane. The total mass of a DMR-type miniature TMD is therefore about 1.5 g, while the oscillating mass is only about 1.0 g. The mass density, Poisson's ratio, Young's modulus, and pre-stress of the membrane were 980 kg/m$^3$, 0.49, 5×10$^5$, and 0.4 MPa, respectively. The pre-stress was estimated by matching the theoretical eigen-frequency obtained from simulations to the experimental one near 100 Hz. The method to measure the dynamic effective mass of the DMR's reported earlier [42] was used here.

In simple terms, the dynamic mass of the DMR was obtained by the measured force acting on the frame divided by the acceleration of the frame. Vacuum grease was applied on the membrane to reduce its Q-factor from around 100 for a pristine DMR to about 25. The second resonance of the DMR's was above 500 Hz so it will not be considered further in this work. Small amount of putty was added onto the platelet to fine tune the first resonance frequency of each DMR, so that collectively they would spread around the first mode of the primary structure. Due to the very small mass of the miniature TMD's, the first mode of the primary structure only shifted by about 0.3 Hz when 4 DMR's were attached, one to each of the four corners.

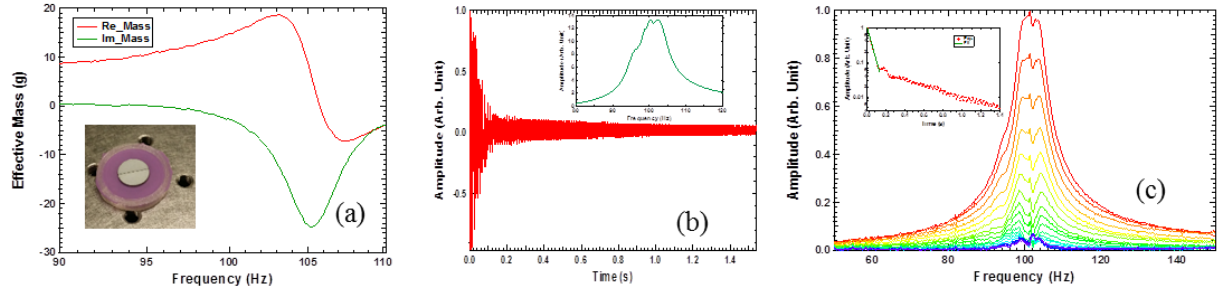

Figure 4, (a) The real part (red curve) and the imaginary part (green curve) of the experimental dynamic mass of a typical DMR damper. The insert is a photo of a DMR damper. (b) The experimental time trace of the displacement of the primary structure with four dampers attached after single impulse excitation. The insert depicts the experimental response function. (c) The delayed Fourier Transform spectra at different delay time obtained from the time trace in (b). The insert is the time dependent peak amplitude obtained from the spectra.

A typical measured dynamic mass of a DMR is shown in Fig. 4(a). The insert shows the photo of a DMR. We chose the convention of negative imaginary mass for energy dissipation and positive imaginary part of the eigen-frequency which then leads to the decay of vibration amplitude with time. At the first resonance of the DMR, the maximum amplitude of the imaginary mass was about 25 g, while the Q-factor obtained from the line width at half height is about 25. It is noted that for a classical TMD, the peak imaginary mass $M_i$ is equal to the Q-factor times the oscillator mass $M_0$, i. e., $M_i = QM_0$, which is well validated here. It is a clear indication that the DMR behaves like a classical TMD as far as its damping effect on the primary structure is concerned. Also, the line shapes of the real and the imaginary parts of the dynamic mass resemble well the Lorentzian form that satisfies the Kramers-Kronig relations.

The measured first resonant frequency of each of the four DMR's is listed in the first row in Tab. 1. These frequencies were the results of fine-tuning so that with these DMR's, one at each corner of the primary structure, the resulting frequency response of the structure is shown in the insert of Fig. 4(b). The line shape of the frequency response nearly resembles a single resonance with an apparent Q-factor of 13 as estimated from the line width, even though the total mass of the dampers is less than 0.1 % of the primary structure, which is in stark contrast to the typical ones in the literature which are more than 10 to 100 times heavier [27, 28, 30 – 32].

Compared to the added mass by multiple TMD's of the order of 5 % or more reported in the literature, our dampers would have been considered as insignificant, let alone causing significant damping. We have also noted that the measured frequency response remained almost unchanged when the damper on the upper left corner was moved to the upper right corner, such that there was no damper on the upper left corner while there were two on the upper right corner.

| Damper # | #1 | #2 | #3 | #4 | #5 | #6 | #7 | #8 |
|---|---|---|---|---|---|---|---|---|
| Group-1 | 99.3 | 99.3 | 105.2 | 105.3 | | | | |
| Group-2 | 96.2 | 98.2 | 100.4 | 101.4 | 101.6 | 101.8 | 102.8 | 104.2 |

Table 1 The measured frequency of the imaginary effective dynamic mass peak of each DMR damper used in the experiments.

The question of whether such frequency response with an 'apparent' Q-factor of 13 would indeed produce a fast decay comparable to such Q-factor is clarified by the experimental time dependent displacement curve shown in Fig. 4(b). Fast decay of vibration is indeed observed within the first 70 ms (~ 7 periods) after an impulse excitation. The remaining slow decay is mostly due to the second eigenmode of the primary structure around 125 Hz. To examine the displacement curve more closely, we carried out Fourier transform of the remaining time-dependent curve after certain delayed time from the impulse, i. e, delayed Fourier transform analysis. The selected moments of time were the ones at the consecutive maximum positive displacement. Each spectrum was therefore taken at about 10 ms (~ 1 period) later than the earlier one. The resulting spectra are shown in Fig. 4(c), where a fast amplitude decrease of the peak around the first resonance of 100 Hz of the primary structure can be clearly seen. In the insert is the amplitude of the peak value versus the delay time. A fast decay in the first stage followed by a second stage slow decay is clearly seen. By fitting the first stage decay curve with an exponent function, the Q-factor of the decay was found to be 14.2, which is in good agreement with the one obtained from the response function of the structure (Fig. 4(b)), confirming the relationship of time trace and response function. The second stage decay was also clearly revealed, with its DTC much smaller than that of the first stage.

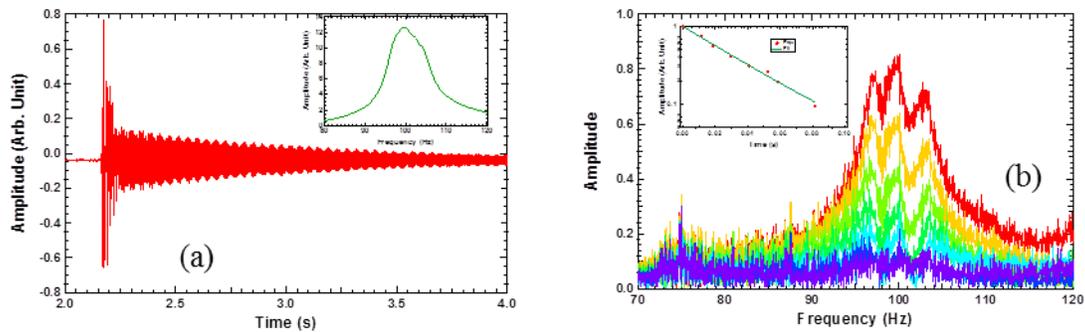

Figure 5, (a) The experimental time trace of the displacement of the primary structure with 8 dampers attached. The insert depicts the experimental response function. (b) The delayed Fourier

Transform spectra at different delay time obtained from the time trace in (a). The insert is the time dependent peak amplitude obtained from the spectra.

To further increase the damping effect, four more DMR's were added and their resonant frequencies were fine-tuned to obtain the experimental response function as shown in the insert of Fig. 5(a), which has an apparent $Q = 11$. The resonant frequencies of the eight dampers are listed in the second row in Tab. 1. The delayed Fourier transform spectra of the decay curve are shown in Fig. 5(b), while the peak amplitude decay curve is shown as the insert in the same figure. The obtained Q-factor is 10.8, which again matches well the one from the response function. It also demonstrated the damping efficiency of the DMR's with total mass of 12 g, or less than 0.2 % of the primary structure, that can bring out a damping ratio for the first eigenmode of the primary structure to be as high as 4.6 %.

**Simulations**

The finite element simulations for the four-damper case in the above experimental investigation were performed by using COMSOL multi-physics with the following materials parameters for the steel plate, density 7850 kg/m$^3$, Young's modulus $2.0 \times 10^{11}$ Pa, and Poison ratio 0.33. The first five eigenmodes for the bare steel plate were shown in Fig. 6(a). The red color indicates high vibration amplitude, while deep blue indicates near zero vibration. The vibration patterns resemble approximately the expression of $\cos(mx) \cdot \cos(ny)$. For damping simulations, instead of calculating the primary structure with the real DMR's, which could take up large amount of computer memory and computational power because of the thin membrane structure of the DMR's, the measured dynamic mass spectra of the DMR's were used instead as the effective input parameters for their damping effects [42]. The dampers were placed at the corners of the plate because it is where the vibration amplitude of the first eigenmode is the largest.

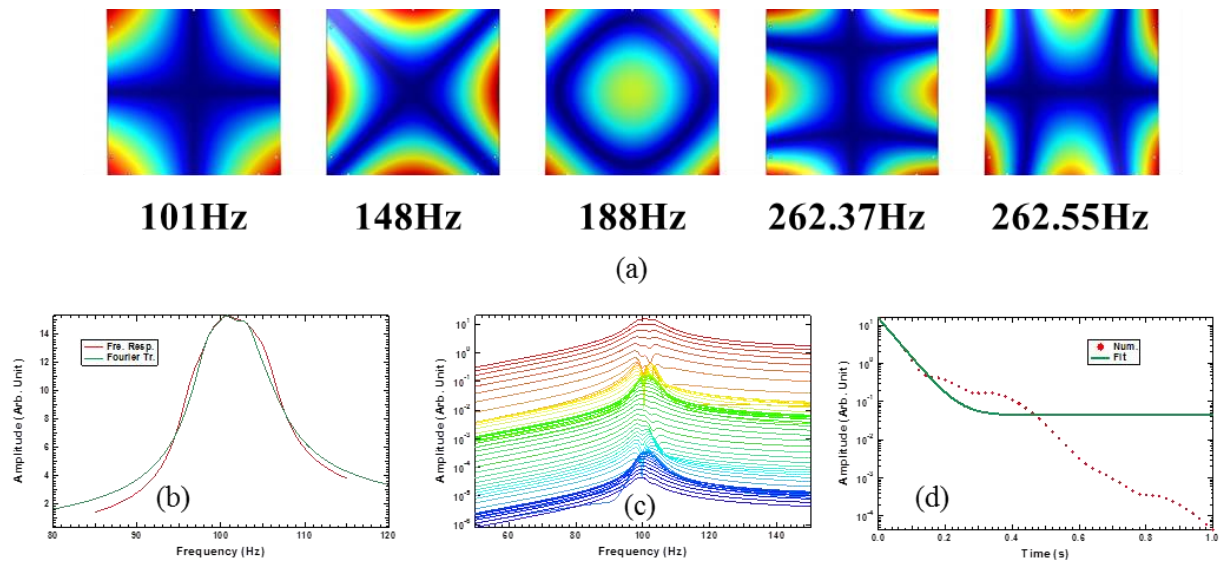

Figure 6, (a) The first five resonant modes obtained from numerical simulations. (b) The frequency response functions of the primary structure with four dampers attached (red curve) obtained directly from simulations, and from mimicking using the five eigenmodes (green curve). (c) The delayed Fourier Transform spectra obtained from the time dependent displacement obtained from simulations. (d) The amplitude of the peak in (c) as a function of delay time (red dashed curve) and the numerical fit by a single exponent function with a small offset (green curve).

The numerical simulations were carried out in the following steps. In step-1, the response function of the primary structure in the same excitation-response scheme as the experiments was calculated when four dampers with the parameters shown in Tab. 1 were attached to the primary structure. The resulting response function is shown as the red curve in Fig. 6(b), with an apparent $Q = 11$. The reason for the discrepancy from the experimental value of $Q = 13$ is most likely due to the imperfection of the real DMR dampers. In step-2 the eigenmodes of the primary structure with the four attached dampers were calculated. The original eigenmode near 100 Hz now splits into five modes, because four additional virtual degrees of freedom were introduced by the four effective TMD's. The eigen-frequencies are {96.4678 + 1.74668$i$, 98.2078 + 2.06091$i$, 100.159 + 1.80042$i$, 102.455 + 2.4188$i$, 104.085 + 2.02823$i$} in Hz, revealing the eigenstates with frequencies spreading across the first resonance of the primary structure and the imaginary parts due to DMR damping. The Q-factor of these resonances ranges from 55 to 43, none being close to the experimental value of 13 under the same situation, or the apparent Q-factor of 11 in simulations. Therefore, the response function obtained in step-1 is a 'mimicked' one by the combination of these resonances. In step-3, the amplitudes of the five eigenmodes were then determined by mimicking the resulting response function following Eq. (2) to that obtained in step-1, assuming that only the eigenmodes around 100 Hz are excited. The resulting response function is shown as the green curve in Fig. 6(b). In step-4, the time dependent displacement was then calculated using Eq. (1) and the outcome from step-2 and step-3. In step-5 the delayed Fourier on the time dependent displacement obtained in step-4 was performed. The resulting spectra are shown in Fig. 6(c). The time interval between two consecutive spectra is about 20 ms, or 2 periods. The evolution of the spectra resembles well the experimental one in Fig. 4(c). Finally, in step-6 the total energy decay curve was calculated by using Eq. (4) and the time dependent displacement found in step-4. The outcome is shown in Fig. 6(d). The decay curve exhibits clearly the characteristic multi-stage fast and slow decays similar to that in Fig. 3. The decay started with a fast drop that lasts for about 100 ms, followed by a slow decay till about 200 ms, then fast decay again. The Q-factor obtained from the first stage fast decay is 12, which is in good agreement with the apparent value of 11 obtained from the spectrum in Fig. 6(b). The decay in the remaining stages is negligible because the vibration amplitude is already minute. Overall, the results from the simulations agree very well with the experimental ones, and are in perfect alignment with the coherent super decay phenomemum.

**Discussions**

After nearly 100 years since the invention of TMD's, the parameter space for optimizing the performance of TMD's seem to have already been exhaustively combed. What we report in this work reveals a piece of fertile and yet unexplored land in the parameter space, where TMD's with mass only a fraction of a percent of the primary structure can effectively tame the individual resonant modes of the primary structure. The intuitive method of using apparent line width in the frequency response function to adjust the design parameters of the TMD's for effective damping is simple to use in designs and in experiments. The non-uniform multiple-stage decay processes where the decay time constant changes in different stages, such that at the beginning there could be a fast decay of most vibration energy, followed by a period of almost stagnating decay, could be very beneficial for practical applications, where the real effect of the remaining few percent of the initial vibration energy on the primary structure after a stage-1 rapid decay is almost negligible. Therefore, the effectiveness of the dampers is almost entirely determined by the first stage.

Although we have only reported the damping of the first resonance of the primary structure, it is straight forward to damp the second resonance which has maximum vibration at the center of each edge (Fig. 6(a)), away from that of the first resonance. One would expect that four dampers similar to the one in Fig. 2(a) are needed. The total added mass to damp the first two resonances would most likely be below 0.5 %. As in most cases the vibration energy is concentrated in the first few resonances of a structure, such approach could suppress the vibration of nearly free bodies with no more than 1% added mass by TMD's.

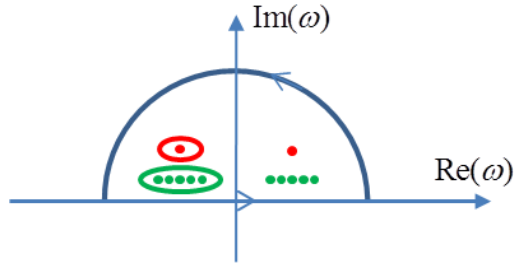

Figure 7, The path (semicircle) to carry out the path integral in Eq. (5), and the simple poles (red and green dots).

Our findings also reveal some interesting underline connections between two complex functions which are mutual approximation of each other. Take the example of the mother resonance $X_M(\omega) = \dfrac{1}{\omega_0^2 - \omega^2 + i\omega\omega_0/Q_0}$ being approximated by five daughter resonances $X_D(\omega) = \sum_{n=1}^{5} \dfrac{A_n}{\omega_n^2 - \omega^2 + i\omega\omega_n/Q_n}$. $X_M(\omega)$ and $X_D(\omega)$ are a mimicking pair, that is, $\operatorname{Re}(X_M) \approx \operatorname{Re}(X_D)$ and $\operatorname{Im}(X_M) \approx \operatorname{Im}(X_D)$, and are expressed in the form of $X_M(\omega) \triangleq X_D(\omega)$ to denote such relation. Furthermore, $x_M(t) \approx x_D(t)$, where

$$x_M(t) \equiv \text{Re}(\tilde{x}_M(t)) = \text{Re}\left(\frac{1}{2\pi}\oint X_M(\omega)e^{i\omega t}d\omega\right) \qquad \text{Eq. (5A)},$$

and

$$x_D(t) \equiv \text{Re}(\tilde{x}_D(t)) = \text{Re}\left(\frac{1}{2\pi}\oint X_D(\omega)e^{i\omega t}d\omega\right) \qquad \text{Eq. (5B)}.$$

The closed integral path is an infinitely large semicircle shown in Fig. 7, with its straight edge along the real axis of the complex $\omega$- plane [41]. The expression $x_M(t) \approx x_D(t)$ means that the two functions are nearly equal in the first stage, as exemplified in Fig. 3 and in Fig. 6(d), and the difference between them is negligible when compared to their initial strength. However, the two complex functions cannot be equal in the domain of complex analysis [41]. The complex function $X_M(\omega)$ has two simple poles with residue '1' in the complex $\omega$-plane, as represented by the two red dots in Fig. 7. The complex function $X_D(\omega)$, on the other hand, has ten simple poles (the green dots in Fig. 7) with proper residues, both taking the values according to the mimicking conditions given in Eq. (1). They are closer to the real axis because they have larger Q-factors and therefore smaller imaginary parts as compared to $X_M(\omega)$. If one carries out the path integral along the red loop in Fig. 7, the integral with $X_D(\omega)$ will be zero while that with $X_M(\omega)$ will be of non-zero. Likewise, along the green loop the path integral with $X_M(\omega)$ will be zero while that with $X_D(\omega)$ is non-zero. In more general terms, consider two complex functions $F_1(\omega)$ and $F_2(\omega)$, then $F_1(\omega) \triangleq F_2(\omega)$ when $F_1(\omega)$ only has two simple poles as $X_M(\omega)$ with the same residues, and $F_2(\omega)$ only has ten simple poles and the residues as those in $X_D(\omega)$, as long as the integrals along the big semicircle is negligible. As any response function can be mimicked by almost an infinite number of combinations of mimicking functions with their own simple poles and residues, a complex function with two simple poles would somehow be connected to many other complex functions with the right poles and residues. Such intriguing mathematics is yet to be explored.

### Acknowledgement

This work was supported by AoE/P-02/12 from the Research Grant Council of the Hong Kong SAR government.

### References

[1]   Abbott, Benjamin P.; *et al.* (LIGO Scientific Collaboration and Virgo Collaboration) Observation of Gravitational Waves from a Binary Black Hole Merger. *Phys. Rev. Lett.* **116** (6): 061102 (2016).

[2]  Weisberg, J. M.; Nice, D. J.; Taylor, J. H. Timing Measurements of the Relativistic Binary Pulsar PSR B1913+16. *Astrophysical Journal*. **722** (2): 1030–1034 (2010).

[3]  Herbert Goldstein, Charles P. Poole, and John Safko, *Classical Mechanics*, 3rd Edition, Pearson (2011)

[4]  L. Cremer, M. Heckl, E. E. Ungar, *Structure-borne sound*, Springer (1988).

[5]  *Modal Testing, Theory, Practice and Application*, 2nd Edition, D. J. Ewings, Research Studies Press Ltd., Baldock, Hertfordshire, England (2000).

[6]  Frahm H. Device for damping vibrations of bodies, US Patent 989958; 1909.

[7]  D. J. Mead, *Passive vibration control*, Willey (1998).

[8]  Housner, G., Bergman, L. A., Caughey, T. K., Chassiakos, A. G. Structural control: past, present, and future. *J. Eng. Mech*. **123**(9), 897–971 (1997)

[9]  Mohamad S. Qatu, Rani Warsi Sullivan, Wenchao Wang, Recent research advances on the dynamic analysis of composite shells: 2000–2009, *Composite Structures* **93** 14–31 (2010)

[10] Spencer, B. F., Nagarajaiah, S. State of the art of structural control *J. Struct. Eng*. **129**(7), 845–856 (2003)

[11] Ormondroyd J., Den Hartog J. P. The theory of dynamic vibration absorber. *Trans. Am. Soc. Mech. Eng*. 50:9–22 (1928).

[12] Optimum damper parameters in terms of frequency ratio and damping ratio, Den Hartog J. P. Mechanical vibrations. 3rd ed. New York: McGraw-Hill (1947).

[13] M. Strasberg and D. Feit, Vibration damping of large structures induced by attached small resonant structures, *J. Acoust. Soc. Am*. **99** (1), 335, (1996)

[14] J. A. Zapfe, G. A. Lesieutre, Broadband vibration damping using highly distributed tuned mass absorbers, *AIAA Journal* **35** (4), 753-755 (1997).

[15] M. N. Hadi, Y. Arfiadi, Optimum design of absorber for MDOF structures, *Journal of Structural Engineering* **124** (11) 1272–1280 (1998).

[16] K. Nagaya and L. Li, Control of sound noise radiated from a plate using dynamic absorbers under the optimization by neural network, *Journal of Sound and Vibration* **208,** 289 (1997)

[17] L. Zuo, S. A. Nayfeh The two-degree-of-freedom tuned-mass damper for suppression of single-mode vibration under random and harmonic excitation, *Journal of Vibration and Acoustics* **128,** 56–65 (2006).

[18] M. S. Khun, H. P. Lee, S. P. Lim, Structural intensity in plates with multiple discrete and distributed spring-dashpot systems, *Journal of Sound and Vibration* **276** (2004) 627–648.

[19] M. B. Ozer, Extending den Hartog's vibration absorber technique to multi-degree-of-freedom systems, *Journal of Vibration and Acoustics* **127** 341–350 (2005).

[20]  R. L. Harne, C.R. Fuller, Modeling of a passive distributed vibration control device using a superposition technique, *Journal of Sound and Vibration* **331,** 1859-1869 (2012).

[21]  Y. L. Cheung, W. O. Wong, $H_\infty$ and $H_2$ optimizations of a dynamic vibration absorber for suppressing vibrations in plates, *Journal of Sound and Vibration* **320,** 29 – 43 (2009).

[22]  L. Zuo, S. Nayfeh, Design of passive mechanical systems via decentralized control techniques, in 43rd AIAA/ASME/ASCE/AHS/ASC Structures, Structural Dynamics, and Materials Conference, 2002, AIAA 2002–1282. H2 and $H_\infty$ optimization

[23]  D. I. G. Jones, Response and damping of a simple beam with tuned damper, *Journal of the Acoustical Society of America* 42 (1), 50–53 (1967).

[24]  H. N. Ozguven, B. Candir, Suppressing the first and second resonances of beams by dynamic vibration absorbers, *Journal of Sound and Vibration* **111** (3) 377–390 (1986).

[25]  R. G. JACQUOT, Suppression of random vibration in plates using vibration absorbers *Journal of Sound and Vibration* **248**(4), 596(2001)

[26]  Paolo Gardonio and Michele Zilletti, L. C. Integrated tuned vibration absorbers: A theoretical study *J. Acoust. Soc. Am*. **134**, (2013)

[27]  Reduced-order multiple tuned mass damper optimization: A bounded real lemma for descriptor systems approach, M. Jokic, M.Stegic, M.Butkovic, *Journal of Sound and Vibration* **330**, 5259–5268 (2011). 1 square meter 4 mm thick steel plate (3.12 Kg) simply supported at four corners, total TMD mass 1.256 Kg, damping factors ~ 0.05 estimated from the FR figures.

[28]  J. Michielsen, I. Lopez Arteaga, H. Nijmeijer, LQR-based optimization of multiple tuned resonators for plate sound radiation reduction, *Journal of Sound and Vibration* 363 166–180 (2016) 0.59 Kg Al plate, up to 35% of added mass of multiple TMD.

[29]  Qi Qin and Mei-Ping Sheng, Analyses of multi-bandgap property of a locally resonant plate composed of periodic resonant subsystems, *International Journal of Modern Physics* **B32,** 1850269 (2018)

[30]  Yiqing Yang, Wei Dai, Qiang Liu Design and implementation of two-degree-of-freedom tuned mass damper in milling vibration mitigation Journal of Sound and Vibration 335 78–88 (2015), 3% of 8 kg = 72 g

[31]  L. Zuo, S.A. Nayfeh Minimax optimization of multi-degree-of-freedom tuned-mass dampers Journal of Sound and Vibration 272 (2004) 893–908, two dampers, each is 3.345 kg and 19.8 cm x 8.382 cm x 2.54 cm

[32]  Jae-Sung Bae, Jai-Hyuk Hwang, Dong-Gi Kwag, Jeanho Park, and Daniel J. Inman, Vibration Suppression of a Large Beam Structure Using Tuned Mass Damper and Eddy Current Damping, *Shock and Vibration* 2014, Article ID 893914, 26 kg eddy current damper

[33] C. R. Fuller, S. J. Elliott, P. A. Nelson, *Active control of Vibration*, Academic Press (1996).

[34] A. Preumont, *Vibration Control of Active Structures: An Introduction*, Kluwer Academic Publishers (2002).

[35] D. Karnopp, M. J. Crosby, R.A. Harwood, Vibration Control using semi-active force generators, *ASME Journal of Engineering for Industry* **96** (2) 619-626 (1974).

[36] C. Maurini, dell'Isola, F. & Del, D. Vescovo, Comparison of piezoelectronic networks acting as distributed vibration absorbers, *Mechanical Systems and Signal Processing* **18,** 1243-1271 (2004).

[37] S-M. Kim, S. Wang, M. J. Brennan, Dynamic analysis and optimal design of a passive and an active piezo-electrical dynamic vibration absorber, *Journal of Sound and Vibration* **330** (4) 603–614 (2011).

[38] R. L. Harne, On the linear elastic, isotropic modeling of poroelastic distributed vibration absorbers at low frequencies, *Journal of Sound and Vibration* **332** 3646-3654 (2013).

[39] Giovanni Ferrari, Marco Amabili, Active vibration control of a sandwich plate by non-collocated positive position feedback *Journal of Sound and Vibration* **342** 44–56 (2015).

[40] Ehsan Omidi, S. Nima Mahmoodi, W. Steve Shepard Jr. Multi positive feedback control method for active vibration suppression in flexible structures Mechatronics 33, 23–33 (2016) .

[41] Mary L Boas, *Mathematical methods in the physical sciences,* 3$^{rd}$ Ed., Wiley (2006)

[42] Liang Sun, Ka Yan Au-Yeung, Min Yang, Suet To Tang, Zhiyu Yang, and Ping Sheng. Membrane-type resonator as an effective miniaturized tuned vibration mass damper. *AIP Advances*, **6**(8):085212 (2016).

[43] Z. Yang, M. Yang, N. H. Chan, and P. Sheng, Membrane-type acoustic metamaterial with negative dynamic mass, *Phys. Rev. Lett.* **101**, 204301 (2008)